\documentclass{article}
\usepackage{spconf,amsmath,graphicx,url,xcolor}
\usepackage{enumitem}

\def\dprime{{$\mathit{d}'$}}
\def\lwlrap{{$l\omega lrap$}}

\title{The benefit of temporally-strong labels in audio event classification}

\makeatletter
\def\@name{ \emph{Shawn Hershey, Daniel P W Ellis, Eduardo Fonseca, Aren Jansen, } \\
    \emph{Caroline Liu, R Channing Moore, Manoj Plakal}\vspace{6pt}}
\makeatother
\address{Google Research, New York, NY, and Mountain View, CA, USA\\
{\footnotesize \tt\{shershey,dpwe,efonseca,arenjansen,carolineliu,channingmoore,plakal\}@google.com}}

\begin{document}
\ninept
\maketitle
\begin{abstract}
To reveal the importance of temporal precision in ground truth audio event labels, we collected precise ($\sim$0.1~sec resolution) ``strong'' labels for a portion of the AudioSet dataset.  
We devised a temporally-strong evaluation set (including explicit negatives of varying difficulty) and a small strong-labeled training subset of 67k clips (compared to the original dataset's 1.8M clips labeled at 10~sec resolution).  We show that fine-tuning with a mix of weak- and strongly-labeled data can substantially improve classifier performance, even when evaluated using only the original weak labels.  For a ResNet-50 architecture, \dprime{} on the strong evaluation data including explicit negatives improves from 1.13 to 1.39. The new labels are available as an update to AudioSet.  
\end{abstract}
\section{Introduction}
\label{sec:intro}

Deep learning classifiers can achieve astonishing accuracies but rely on large amounts of training data.  For sound event classification, this data generally has to be directly and expensively gathered from annotators (instead of, say, inferred from existing information).  The fastest approach is collecting temporally-imprecise annotations (e.g., the annotator indicates if a sound event is present within a 10~sec clip, but does not provide more detailed timing), and thus most of the large-scale classifiers developed so far have used such annotations \cite{audioset}, which we will here refer to as ``weak''.  

This raises the question of the extent to which accuracy is impaired by the weakness of the labels.  To answer this, we collected temporally-precise (hereafter, ``strong'') labels via an annotation interface in which annotators indicated precise time extents for each marked event on a spectrogram. We collected about 81k such annotations, of which 14k correspond to our existing evaluation set (excluding clips that have been deleted since the original release\footnote{Since defining the AudioSet segments in March 2017, around 13\% have become unavailable, an attrition rate of $\sim$0.3\% per month.}), leaving up to 67k for training. 

This represents only about 4\% of the 1.8M training clips for which weak labels were collected, so it is doubtful that the improved precision of the labels will compensate for the scarcity of data if used alone.  However, we may hope that combining a large, weakly-labeled dataset with a small, strongly-labeled subset can deliver a classifier superior to one trained on either dataset individually. 

\begin{figure}[t!]
\centering
\includegraphics[width=8.5cm]{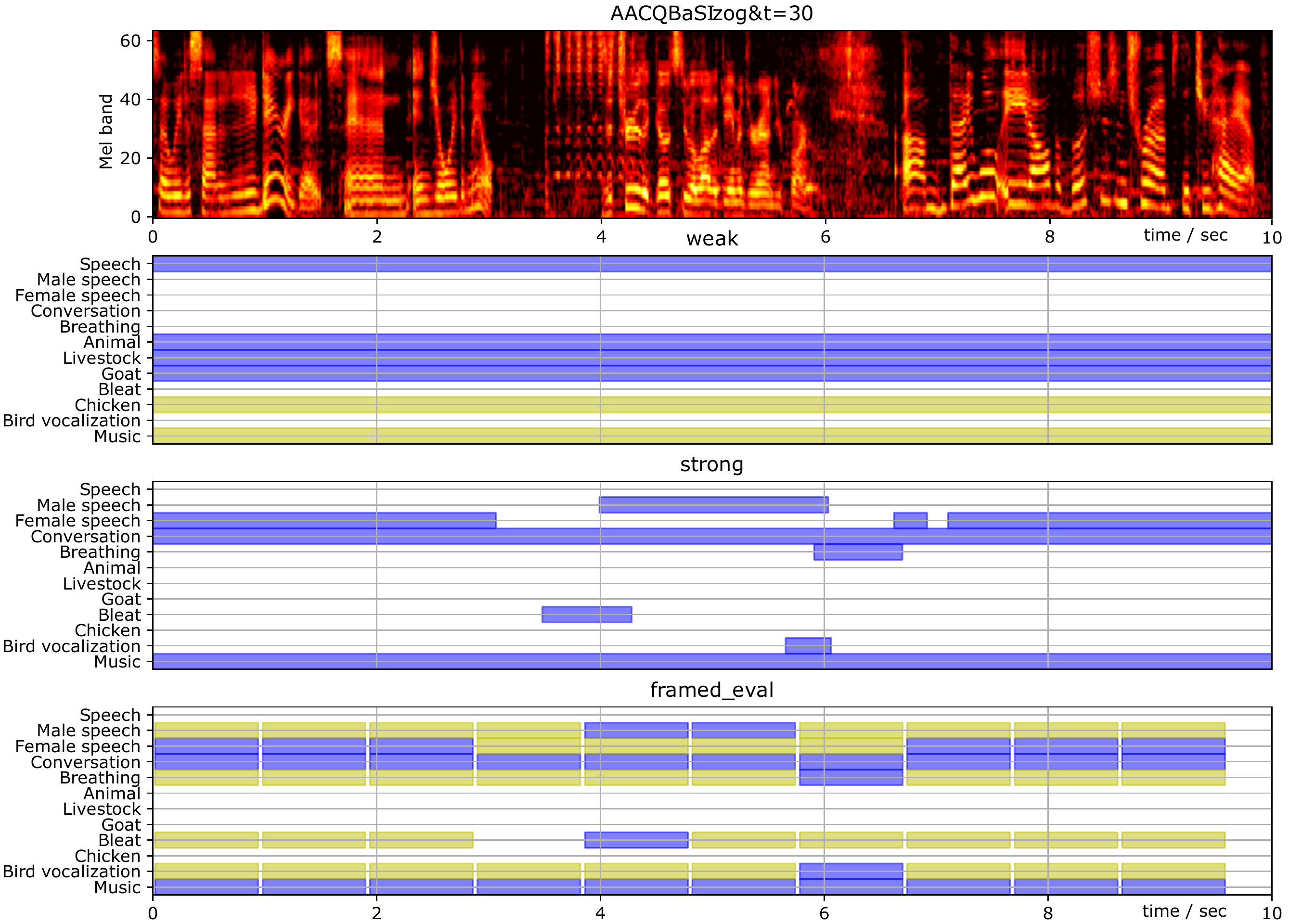}
\caption{Comparing the spectrogram of one clip (top row), the original 10~sec-resolution labels (second row, blue indicates present, yellow indicates confirmed not present), the temporally-strong labels collected in this work (third row), and the strong labels projected onto 0.96~sec frames along with ``complementary negatives'' used for strong-label evaluation (see section \ref{data:evaluation}; gaps for clarity only).}
\label{fig:strong_labels_example}
\end{figure}

\section{Related Work}
\label{sec:related}
Datasets incorporating strong labels appeared with the first editions of the DCASE Challenge \cite{mesaros2017detection}. 
Because manual annotation of sound events’ start and end times is time-consuming, the datasets are limited in size. For example, the TUT Sound events 2016 \cite{Mesaros2016_EUSIPCO} and TUT Sound events 2017 \cite{DCASE2017challenge} each total $\sim$2~h of labeled audio. These are among the few examples of strongly-labeled real-world audio training sets.
The dataset released for DCASE 2017 Task 4 (``Large-scale weakly supervised sound event detection for smart cars’’ \cite{DCASE2017challenge}) only provides strong labels for the validation and test sets (totalling less than 5~h), comprising data from 17 AudioSet \cite{audioset} classes.

Synthetic datasets, in which soundscapes are generated programmatically by mixing a set of target sound events and background audio, have become popular since they allow precise event times without the expense of manual labeling.  Additional advantages include precise control of event amplitudes, and potentially unlimited training set sizes (albeit with limited diversity).
The main shortcoming is that synthetic soundscapes may not always be representative of real-world conditions \cite{salamon2017scaper}.
Examples include URBAN-SED \cite{salamon2017scaper} which is synthesized by mixing sound events from the 10 classes of UrbanSound8K \cite{salamon2014dataset} using the Scaper library. 
DESED \cite{Turpault2019} has been used for DCASE Task 4 in recent years, covering 10 classes
of domestic sounds. It features several data subsets of recorded and synthetic soundscapes, where human-provided strong labels are reserved for evaluation purposes. In addition, code is provided for synthetically generating new soundscapes.
None of these datasets feature more than 20 classes.

Leveraging only weakly labeled data to train audio event recognizers has been extensively investigated \cite{PANN,autopool,fonseca2020addressing,hershey2017cnn,ford2019deep}.
For instance, there has been substantial progress made since the AudioSet data were released in 2017.  Current state-of-the-art systems include effective techniques to mitigate the large class imbalance and benefit from data augmentation \cite{PANN}.
Other works use strategies including pooling \cite{autopool} and attention \cite{ford2019deep} to train on weak labels while still producing scores at fine time resolutions. Evaluating these improvements has been difficult in the absence of strongly-labeled test data.

The labels used in this paper amount to a total of over 200~h of audio with both strong positive and explicit negative labels across 356 classes from the AudioSet ontology (see section \ref{sec:datarelease} for more details).
This unprecedented volume of strong labels can support the investigation of combining strong and weak labels to train audio event classifiers.
Research on how to exploit the combination of both strong and weak labels is scarce;
one of the few previous works following this trend adopts an approach based on manifold regularization on graphs \cite{kumar2017audio}.
A performance boost is reported by adding a small amount of strongly labeled data to the weak labels, possibly due to the mitigation of the inherent problems in weakly labeled data  \cite{kumar2017audio,Fonseca2019learning,turpault2020limitations}.
In this work, we also show substantial benefits when combining both types of labels in the context of AudioSet.

\section{Strong-Labeled Dataset}
\label{sec:dataset}

The original AudioSet data \cite{audiosetwebsite} comprised a set of around 2M 10~sec excerpts from video soundtracks, each bearing an average of $\sim$2 labels drawn from a 527-entry sound ontology; the total number of examples of each label ranged from around 100 (for ``Toothbrush'') to around 1M (for ``Speech'').  The annotation process for each excerpt involved human annotators confirming or refuting the presence of a small set of labels (automatically proposed via metadata or audio similarity).  Annotators judged between 2 and 15 labels per clip.

This approach resulted in several limitations.  Firstly, the temporal precision of the labels is limited to the 10~sec duration of the clip (``weak labels'').  In many cases, the sound event in question will in fact occupy only a small portion of the clip.  Secondly, only a small subset of labels is confirmed (as either present or affirmatively absent) for each clip; there are likely many instances of sound events present in the clip about which the annotators were never asked, and which therefore are not reflected in the labels (``missing positives'').  Thirdly, some labels may be inappropriately abstract:  AudioSet labels are arranged in a nearly-strict hierarchy (apart from some cross-links), with a parent node intended for use only for sounds that do not belong in any of the child nodes.  For instance, the ``Snake'' node has two subnodes, ``Hiss'' and ``Rattle'', but some snakes may make sounds that fall into neither of these categories. Only these otherwise-unassignable sounds should be labeled with the ``Snake'' parent node.  However, during label confirmation an annotator might be shown a snake hissing, but offered only the more abstract ``Snake'' label to confirm, and would naturally rate ``Snake'' as present, whereas strictly the label would have been only ``Hiss''.

The new label collection aimed to resolve all three of these issues. Instead of simple present/not present checkboxes, the annotators interacted with multiple timelines alongside a spectrogram representation, on which they could drag out time regions indicating the extent of each sound event (comparable to Audio Annotator \cite{cartwright2017seeing}).  Annotators quickly became adept at marking time regions on this display and we judge their timings to be precise at least to 0.1~sec resolution (based on informal spot-checks). Annotators then assigned the appropriate label from a searchable pop-up list that would show all available labels, including children, enabling them to follow the instructions to choose the single most-specific label for each sound.  (The annotators apparently became familiar with the entire lexicon of available labels, judging from the largely complete coverage of their responses and spot-checks.)  Finally, the annotators were instructed to mark every sound event they perceived in the clip, eliminating the problem of missing positives. 

We made a selection from the original AudioSet clips to be labeled under the new scheme.  We included all the 18k surviving evaluation clips, and additionally selected around 67k clips from the 1.8M available training clips.  Rather than selecting at random, we attempted to include 250 from each class (some classes had fewer than this in total, in which case every available example was used). However, label co-occurrence meant that many (weak) labels (including the most common labels, ``Speech'' and ``Music'') appeared far more often in the data sent for relabeling.

For expediency, we decided to subsume all the musical instrument and music genre labels from the original AudioSet into the single ``Music'' label.  To identify and mark every instrument in ensemble recordings is extremely arduous, and in general these labels behave very differently from environmental sound events.  This eliminated 140 of the original 527 label categories.

To improve label accuracy, we used a ``pipelined'' process, where a first-pass labeling was reviewed by a different annotator who could modify the labels, and we accepted labels once they passed such a stage without changes.  Even allowing for 5 stages, however, this rarely converged to complete consensus, so eventually we settled for a single stage of review (i.e., each labeling is the result of two annotators' input). We have not estimated inter-annotator agreement, but are satisfied with the labels after spot-checking.


\subsection{Strong Labeling Analysis}
\label{data:results}

\begin{figure}[t]
\centering
\includegraphics[width=8.5cm]{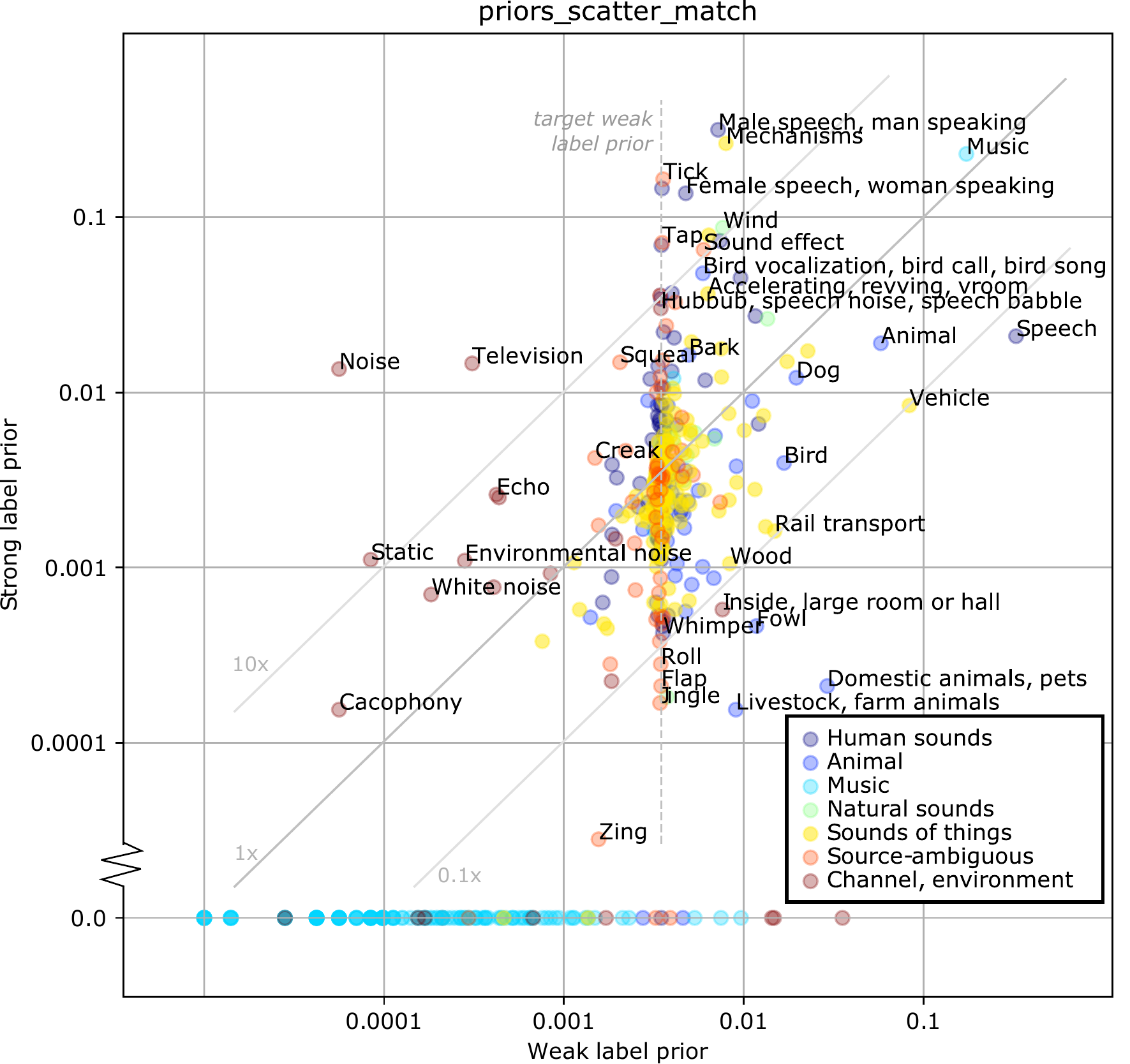}
\caption{Scatter plot of per-class priors for all 527 classes for original (x-axis) and strong (y-axis) labels.  Note that 151 labels, including all the children of ``Music'', were given no strong labels, and appear at the bottom of the graph with a strong label prior of 0.}
\label{fig:priors_scatter}
\end{figure}


Figure \ref{fig:priors_scatter} plots the scatter of overall incidence (priors) for weak versus strong labeling for all 527 classes, including the 151 for which no strong labels were collected (strong label prior $=$ 0).  Diagonal lines show the locus for classes whose priors are unchanged, or changed by 10$\times$ and 0.1$\times$.  
``Speech'' and ``Vehicle'' are among the classes whose priors have dropped below 0.1$\times$, whereas ``Male speech'', ``Mechanisms'', and ``Tick'' have grown more than 10$\times$.  We can also see the cluster around the target weak label prior $\approx 0.0038$ (250 in 67k) revealing the relabeling selection policy.

These results provide evidence that the annotators were successful in fulfilling their instructions to label every salient sound event, since there are almost twice as many positive class/clip instances after strong labeling -- 217k versus 147k.  However, the large changes in priors evident in Figure \ref{fig:priors_scatter} indicate that the (effective) interpretation of many labels has changed significantly between the label sets.  Because each clip represents multiple classes (3.5 on average in the strong labels, 2.4 in the weak), we cannot simply make correspondences between strong labels and weak labels -- since the strong labels may relate to other weak labels present in the same clips.  We can, however, calculate the {\em odds ratio} (OR, the change in the ratio of positive to negative outcomes due to the condition~\cite{szumilas2010explaining}) as a measure of the influence of a label in one set on the likelihood of label presence in the other set.  For instance, the strong labels with the largest odds ratio related to the weak ``Speech'' label are ``Male speech'' (OR=12.1), ``Female speech'' (OR=8.9), ``Children playing'' (OR=8.8), and ``Conversation'' (OR=8.7); the strong ``Speech'' label receives an OR of only 2.7.  Looking the other way, we can find the weak labels with the largest relationship to the strong label ``Mechanism'': they are ``Mains hum'' (OR=6.5), ``Gears'' (OR=5.2), ``Cupboard'' (OR=5.1), ``Cutlery'' (OR=4.8), ``Hammer'' (OR=4.7), and ``Ratchet'' (OR=4.6).  In general, we see shifts in abstraction and preferences for certain labels which underline that the effective interpretation of the ontology has shifted between strong and weak label sets; this label mismatch confers a disadvantage when using the original weak labels to evaluate classifiers trained on the strong labels, and vice-versa.

Note that in our training and evaluation, we use the labels exactly as provided or confirmed by the annotators.  Some of the differences in annotator behavior described above could be addressed by automatic canonicalization of labels.  For instance, although the annotators were instructed to use the single most-specific label for all instances, these labels could be augmented by all parents in the class ontology (ignoring the question of how to handle the few nodes that have multiple parents).  In this way, the ``Snake'' class would automatically be trained to treat all instances of ``Hiss'' and ``Rattle'' as positives.  Similarly in evaluation, test instances of ``Hiss'' would also be evaluated as positives for ``Snake''.  Preliminary investigation of this kind of ``label smearing'' (based on similar ideas from image classification \cite{sun2017revisiting,gao2017knowledge}) did not yield significant changes in metrics, but it warrants further attention.

\subsection{Strong Label Evaluation Set}
\label{data:evaluation}

In addition to the original weak-label evaluation which rated 10~sec clips according to the average classifier score over the whole clip, we defined a new evaluation set based on the strong relabeling of the original AudioSet evaluation data.  Our classifiers operate on 960~ms input patches (chosen to capture enough information to distinguish most events), so we chose to re-frame the variable-duration strong labels onto that grid: a given 960~ms frame inherits a label if it is either at least 50\% filled with the label (i.e., at least 480~ms), or if it contains at least 50\% of the total label duration (so strong label segments shorter than 480~ms total will still be reflected). 

For the original AudioSet data release, we did not include any of explicit negatives that had been collected, i.e., labels where the annotators had explicitly confirmed that a class was not present in the clip.  Because any single class (with a few exceptions) is rare, it is usually correct to assume that any class not labeled a positive is in fact a negative, and as such can be used in evaluation to estimate the performance of the classifier for negative inputs.  The impact of this in evaluation is that a given classifier typically has far more negatives than positives, e.g. the median count of positives in the $\sim$18k clip evaluation set is under 60, meaning there more than 300 times as many negatives as positives.  The majority of these ``implicit negatives'' are from unrelated classes and thus easy to distinguish, leading to possibly misleading performance metrics. More revealing are metrics that focus on the classifiers' ability to distinguish true positives from the most confusable (``hard'') negatives.  

We collect more useful evaluation results in two ways:  Firstly, we only use as negatives clips that have been {\em explicitly} marked as such by annotators, and we solicit annotations of a given class primarily for examples that appear plausible candidates -- in particular, after training a classifier on the original data, we took the clips that resulted in the highest scores for each class under that classifier and had the annotators confirm or refute the presence of those classes.  Many of these were in fact negatives, but they are by construction the negatives that are most challenging for our classifiers, and thus best able to reveal marginal classifier improvements.  Note that restricting evaluation to these ``hard'' negatives presents a much more challenging task for the classifier, so measures such as \dprime{} are much lower than those reported on implicit-negative evaluations.

Secondly, given the availability of strong labels, we particularly wanted to be able to reward classifiers that showed a contrast between regions containing an event and the surrounding context.  For this, we added ``complementary negatives'': if a clip contains some positive frames for a given class, then any surrounding frames that contain less than 50\% of that class's labeled segments are marked as explicit negatives for that class.  Overall evaluation mixes the scores of the ``complementary negatives'' with the ``explicit negatives'' to give balanced measures of performance.

The bottom pane of Figure \ref{fig:strong_labels_example} shows an example of the evaluation labels resulting from these steps.

\subsection{Dataset Release}
\label{sec:datarelease}

We are releasing strong positives and explicit negatives for 356 of the original 527 AudioSet classes (excluding ``Music'' subclasses, along with some other rare classes).  The training set consists of a 66,924 clip subset of the original AudioSet, with between 1 (``Canidae'') and 22,410 (``Male speech'') clips with strong positives per class (mean 418.9, median 248).  Explicit negative labels are provided with between 4 (``Electronic tuner'') and 38,220 (``Music'') clips per class (mean 758.3, median 361).  
From the original AudioSet evaluation set, we provide 14,470 clips with between 14 (``Ping'') and 4774 (``Male speech'') positive clips per class (mean 155.1, median 58), and 32 (``Light engine'') to 216 (``Mosquito'') negative labels (excluding ``complementary negatives'') per class (mean 124.3, median 123).  All strong labels are provided at the full, original time resolution, which is sufficient to derive the framed evaluation data described in section \ref{data:evaluation}.  
The label information appears as alternates to the weak-label CSV files on the AudioSet download page.\footnote{\url{https://research.google.com/audioset/download.html}}

\section{Experiments}
\label{sec:experiments}

\label{sec:framework}

To investigate the benefit of strong labels on classification, we formed several training sets:

\begin{itemize}[itemsep=2pt,parsep=2pt,topsep=2pt,partopsep=2pt]
    \item Weak-1.8M - The original AudioSet training set (minus subsequently deleted videos).
    \item Strong-67k - The new strongly labeled training data.
    \item Weak-67k - Subset of Weak-1.8M containing only the videos present in Strong-67k.
    \item Diffuse-67k - The clips from Strong-67k with each label expanded to the entire 10~sec.  This is comparable to Weak-67k but using the same label interpretations as Strong-67k, as discussed in Section \ref{data:results}.
\end{itemize}

We additionally fine-tuned the model trained on the Weak-1.8M data using mixtures of that data and, separately, Diffuse-67k and Strong-67k.
Mixing is controlled by a real-valued hyperparameter $\mu$ in [0,1] where 0 selects only Weak and 1 selects only Diffuse/Strong.


For all experiments, for each epoch, we selected a random 960~ms frame of the 10~sec clip to produce a 96$\times$64 log-mel spectrogram with the same parameters as described in the original AudioSet work \cite{hershey2017cnn}. Strong labels are applied using the same 50\% overlap principle described in section \ref{data:evaluation}. We trained all models with ResNet-50 \cite{he2016deep} and the Adam optimizer \cite{hershey2017cnn}.



\subsection{Results}

We report two metrics for each evaluation set: \dprime{} \cite{hershey2017cnn} is a measure of the separation of positive and negative examples within a single classifier, calculated from the area under the ROC curve.  \lwlrap{} \cite{fonseca2019audio} indicates the precision of lists of ranked labels obtained for each test clip, and thus, unlike \dprime{}, reflects the relative calibration of score values from different class outputs.

As discussed in \ref{data:evaluation}, \dprime{} results are calculated using only ``explicit negatives''.
This makes the \dprime{} values much smaller than in previous work (since the negatives are, on average, more similar to positives), but also gives a more sensitive metric primarily measuring the improvement on the most difficult examples, rather than being dominated by easy examples. \lwlrap{} depends only on positive labels (specifically, the rank of classes present in a particular clip), so is unaffected by the definition of negative labels in the evaluation set.


The metrics \dprime{} and \lwlrap{} converge to their maxima, in general, at different mixing parameters $\mu$ (in the case of fine-tuning) and at different points during training. We use a held out validation set to pick a single $\mu$ and checkpoint for each model ($\mu = 0.8$ for +Strong and $\mu = 0.7$ for +Diffuse). 

Table \ref{tab:main_results_external} shows the results of training with the various datasets. For the Strong eval, training with the Strong-67k produces the best results among the 67k-clip training sets.  
For these sets, we can separate the effects of label mismatch (Weak-67k versus Diffuse-67k) and weak versus strong labels (Diffuse-67k versus Strong-67k); as expected, Weak-67k is far better than Diffuse-67k on the Weak eval since the labels are matched; Diffuse-67k is better for the Strong eval, at least for \dprime{}; confusingly, it is no better than Weak-67k for \lwlrap{}.  However, adding the strong labels
via Strong-67k gives a healthy improvement of 0.17 \dprime{} on the matching Strong eval, and even improves the Weak eval \dprime{} by 0.14, despite the label mismatch.

Training on the Weak-1.8M examples substantially improves over training on the Strong-67k examples alone. However, when we pre-train with the Weak-1.8M and then fine-tune on a mixture of the Weak-1.8M and Strong-67k labels we get the best results. When fine-tuning, we get a 0.11 \dprime{} improvement on the Strong eval by using Strong instead of Diffuse labels. As discussed previously, there are multiple factors that make the strong labels better. The only difference between +Diffuse and +Strong is the temporal precision, so this suggests that the 0.26 \dprime{} improvement from Weak-1.8M to +Strong can be split between 0.11 for the improved temporal precision and 0.15 for all other factors. 

We again see that Strong labels improve \dprime{} on the Weak eval as well, though by a smaller amount.  \lwlrap{} improves on Strong eval and reduces on Weak eval when using Diffuse labels (reflecting the impact of label interpretation match), with Strong labels give better \lwlrap{} compared to Diffuse in both cases (although worse than no fine-tuning for the Weak eval). Not shown in the table is that, if we were to choose a $\mu$ and checkpoint to optimize each metric separately, we could achieve 1.4 \dprime{} and 0.48 \lwlrap{} on the Strong eval. Typically, $\mu$ fell in the range 0.7 to 0.95 when optimizing \dprime{}, but was around 0.2 to optimize \lwlrap{}.

\begin{figure}
\centering
\includegraphics[width=0.9\columnwidth]{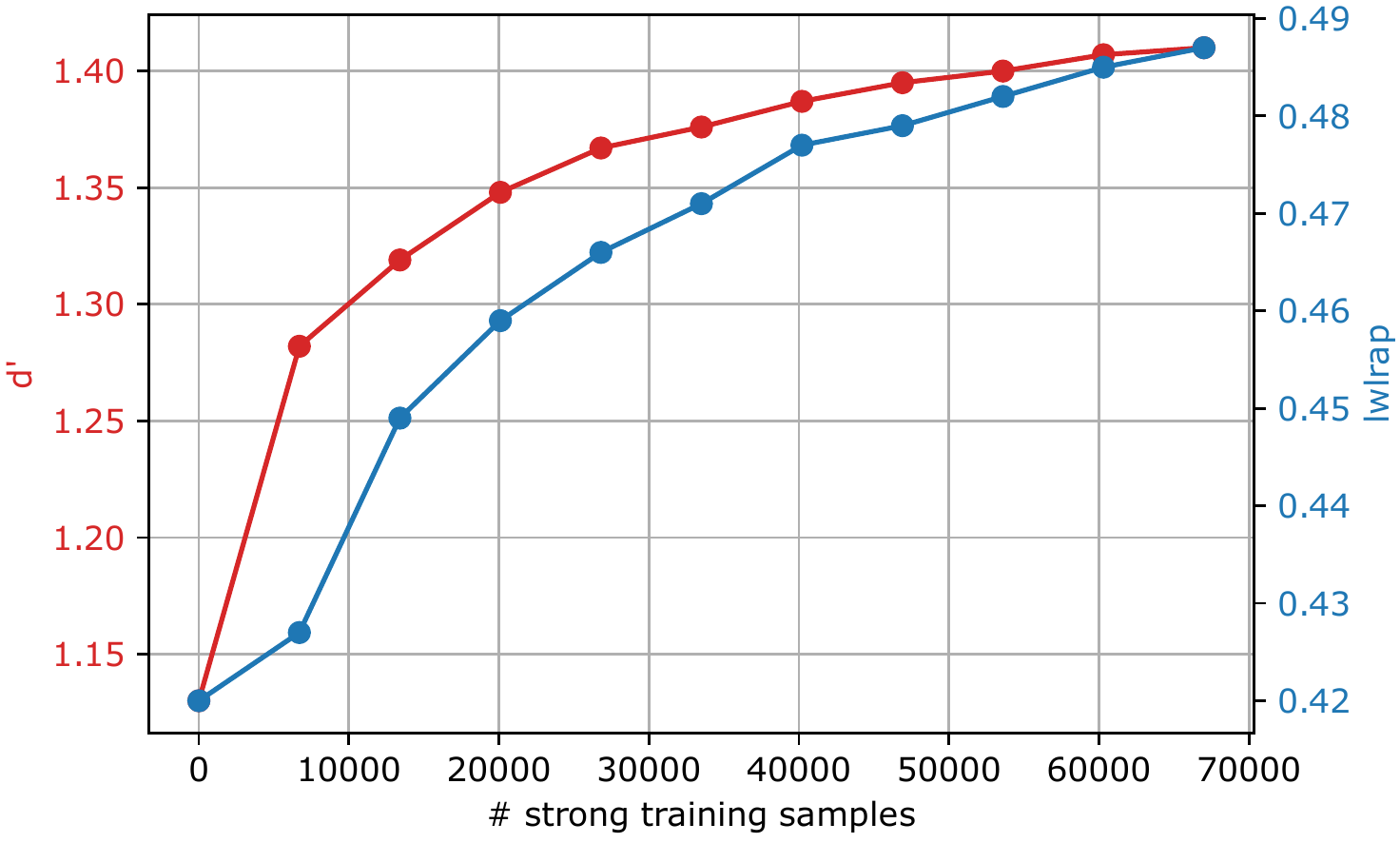}
\caption{\dprime{} and \lwlrap{} on the Strong eval data as a function of fine-tuning with varying proportions of the Strong-67k clips.}
\label{fig:sizez_dprime}
\end{figure}

Figure \ref{fig:sizez_dprime} shows the impact of varying the proportion of the strong training data used.  There is a steady improvement of both metrics as we vary the Strong data from 10\% to 100\%; at 67k clips, we are reaching diminishing returns, at least for this training scheme.


\begin{table}[t]
  \caption{Evaluation results.}
  \vspace{0.1cm}
  \centering
  \begin{tabular}{ l c c c c p{1cm}}
    \hline
     & \multicolumn{2}{c}{Weak eval} 
     & \multicolumn{2}{c}{Strong eval} \\
    Training dataset & \dprime{} & \lwlrap{} & \dprime{} & \lwlrap{} \\ \hline
    Weak-67k & 0.86	& 0.39 & 0.81 & 0.29 \\
    Diffuse-67k & 0.82 & 0.27 & 0.88 & 0.29 \\
    Strong-67k	& 0.96 & 0.31 & 1.05 & 0.33 \\
    \hline
    Weak-1.8M	& 1.17 & \textbf{0.53} & 1.13 
    & 0.43 \\
    \hline    
    + Diffuse & 1.21 &	0.38 & 1.28 & 0.43 \\
    + Strong & \textbf{1.28}    & 0.42	& \textbf{1.39} 
    &  \textbf{0.47} \\
    \hline
  \end{tabular}
  \label{tab:main_results_external}
\end{table}

\section{Conclusions}
\label{sec:conclusions}

We show that employing labels with fine time resolution (``strong labels'') for even a few percent of the training set delivers significant improvements via fine-tuning a classifier previously trained on a large dataset with temporally-weak labels. 
%
Future work includes investigating the interaction of annotator-based strong labels with automatic means for mitigating weak labels such as autopool \cite{autopool}, or related work from vision on using labels of varying quality \cite{noisyData}. 
Our evaluations were on the basis of fixed-size frames, but strong labels suggest approaches that directly predict segment boundaries.

\vfill\pagebreak

\bibliographystyle{IEEEbib}
\bibliography{refs}

\end{document}